\documentclass[preprint,preprintnumbers,amsmath,amssymb,prl]{revtex4}
\usepackage[pdftex]{graphicx}
\usepackage{dcolumn}
\usepackage{bm}
\usepackage{float}
\usepackage{color}
\usepackage{hyperref}
\usepackage{natbib}
\usepackage[T1]{fontenc}
\usepackage{wasysym}
\begin{document}

\title{Electric field induced gelation in aqueous nanoclay suspensions}
\author{Paramesh Gadige}
\author{Ranjini Bandyopadhyay}
\email{ranjini@rri.res.in}
\affiliation{Soft Condensed Matter Group, Raman Research Institute, C. V. Raman Avenue, Sadashivanagar, Bangalore 560 080, INDIA}

\begin{abstract}
Aqueous colloidal Laponite clay suspensions transform spontaneously to a soft solid-like arrested state as its aging or waiting time increases. This article reports the rapid transformation of aqueous Laponite suspensions into soft solids due to the application of a DC electric field. A substantial increase in the speed of solidification at higher electric field strengths is also observed. The electric field is applied across two parallel brass plates immersed in the Laponite suspension. The subsequent solidification that takes place on the surface of the positive electrode is attributed to the dominant negative surface charges on the Laponite particles and the associated electrokinetic phenomena. With increasing electric field strength, a dramatic increase is recorded in the elastic moduli of the samples. These electric field induced Laponite soft solids demonstrate all the typical rheological characteristics of soft glassy materials. They also exhibit a two-step shear melting process similar to that observed in attractive soft glasses. The microstructures of the samples, studied using cryo-scanning electron microscopy (SEM), are seen to consist of percolated network gel-like structures, with the connectivity of the gel network increasing with increasing electric field strengths. In comparison with salt induced gels, the electric field induced gels studied here are mechanically stronger and more stable over longer periods of time.
\end{abstract}

\maketitle

\section{Introduction}
Rheological properties of certain complex fluids or colloidal suspensions can be tuned reversibly by applying small amplitude electric or magnetic fields. Such fluids are known as electrorheological (ER) or magnetorheological (MR) fluids respectively. Understanding the flow behavior and microstructural changes of such fluids due to the application of external fields is of great importance from an academic point of view as well as due to their industrial applications \cite{{Sheng}, {Chin}}. ER fluids serve as electric-mechanical interfaces in sensors and devices and are classified as smart fluids. External fields applied to a colloidal suspension can drive the system from equilibrium to non-equilibrium or vice-versa and are studied in the context of many body statistical mechanics \cite{Lowen}. For example, electric field induced pattern formation was reported in non-aqueous colloidal dispersions of $ BaTiO_{3} $ in castor oil  \cite{{Trau}}. In these experiments, columns, disks and more complex structures were obtained by applying appropriate electric fields. The gelation of natural polymer (cellulose, starch, and chitosan) dispersed suspensions under electric fields has also been reported \cite{Ko}. Clustering of colloidal particles under the influence of an ac electric field and frequency dependent shape changes of clusters were studied in aqueous colloidal dispersions of polystyrene particles \cite{sood}. Furthermore, aqueous suspensions of a disk-shaped natural beidellite clay were reported to reorient into a highly aligned nematic liquid-crystalline phase in the presence of an ac electric field, with the nematic director aligned perpendicular to the electric field \cite{Paineau, Paineau2}.  

Colloidal clay suspensions are investigated owing to the rich sequence of phases that these materials exhibit eg. fluids, gels, ordered and disordered phases \cite{{Luckhum}, {Dijkstra},{Mourchid}, {DBonn98}, {ruzika2}, {Herman}}. The changes between these phases can be brought about by the addition of salt or by increasing the colloidal volume fraction. Laponite is a synthetic smectite clay system which has been studied for its rheological merits and also as a fundamental model system to understand the colloidal glass transition \cite{{saha},{joshi_JCP}, {paramesh}, {Herman}}. The structure of Laponite closely resembles the natural clay mineral hectorite with chemical formula Na$_{+0.7}$[(Si$_{8}$Mg$_{5.5}$Li$_{0.3}$)O$_{20}$(OH)$_{4}$]$^{-0.7}$ and has a discotic shape with $25$ nm diameter and $1$ nm thickness. When Laponite particles are dispersed in aqueous medium, $Na^{+}$ ions from the Laponite discoids dissociate from its surface. This results in a large negative surface charge on the faces. Furthermore, there is a small positive charge on the rim of each Laponite particle if pH<11. The electrostatic interactions that develop between the particles give rise to rich colloidal phases at different Laponite concentrations and sample aging times or waiting times ($ t_{w} $, the time evolved since the sample preparation). Interestingly, Laponite clay aqueous suspensions can form $Wigner$ glasses or repulsive glasses \cite{DBonn98} at concentrations as low as $2-3.5$ wt\%. Several photon-correlation spectroscopy measurements \cite{{saha_EPL},{sahaLagm},{saha}, {ruzica},{saha2}, {Herman}, {ruzika}} and rheology studies \cite{{cates_sgr}, {joshi}, {Mourchid}, {joshi2}} have reported the slowing down of the structural or $ \alpha $-relaxation time as the suspensions age or $ t_{w} $ increases.

The effects of electric field on various physical phenomena in colloidal Laponite clay suspensions and their hydrogel composites have been reported in the literature. Phytagel-Laponite hydrogel composites were found to behave as intelligent hydrogel composites that effectively respond to the stimuli of environmental pH and electric field changes \cite{Ekici}. Laponite colloidal particles were also used to obtain mechanically strong organic-inorganic nanocomposite (NC) gradient hydrogels in a direct-current (DC) electric field due to the directional movement of Laponite clay particles \cite{YTan}. Tarafdar et.al. have reported the effects of an electric field on desiccation crack patterns in drying Laponite gels \cite{tarafdar2007,tarafdar2008, tarafdar2008IECR, tarafdar2013}. Upon the application of radial fields acting inward or outward, radial cracks patterns were observed in the setup when the center terminal was positive, while cross-radial cracks were noticed when the center was held at a negative potential \cite{{tarafdar}}. The dessication cracks that developed due to the applied electric field were seen to be sensitive to the field strength, frequency of the ac field, its direction, and the time of exposure to the electric field (memory effect) \cite{tarafdar2016,tarafdar2017, tarafdar2018, tarafdarL2013, tarafdarIJP}. In addition, ER properties of Laponite clay aggregates dispersed in silicone oil were studied in rheometric studies by applying a DC electric field \cite{parmar}. Chain-like microstructures were formed above a certain amplitude of the triggering electric field and a significant yield stress was measured in these samples. Viscosity changes in aqueous hectorite (Na$_{0.3}$[(Mg$_{2.7}$Li$_{0.3}$)Si$_{4}$O$_{20}$(OH)$_{2}$]) clay suspensions were reported due to the application of DC electric fields \cite{Kimura}. The gradual increase in the viscosity was attributed to the formation of a three-dimensional network structure due to the deformation of the electric double layer as a result of the applied DC electric field. 

In the present article, we report the electric field induced irreversible soft solidification of aqueous Laponite clay suspensions over the surface of the positive electrode. We study the samples that are formed due to the application of increasing electric field strengths by monitoring their rheological behavior. We understand these results in terms of the microscopic structures of the samples after the application of electric fields of varying strengths by directly imaging the sample using cryogenic scanning electron microscopy. Our imaging  study reveals the formation of network-like percolated structures, thereby confirming that the soft solids formed under the applied electric fields have gel-like morphologies.  We believe that our study will trigger further insights in producing strong hydrogels, with implications in the design of soft machines \cite{PCalvert}.

\section{Experimental methods}

Laponite XLG $(R)$ (BYK, Inc.) was dried in a hot air oven at 120$^{\circ} $C for 20 h to remove the moisture absorbed in the sample. An appropriate quantity of the dried Laponite powder was weighed and then dispersed in Millipore water (resistivity $18.2$ M$\Omega$-cm) to prepare a suspension of concentration $ C_{L} $ = 3.0 wt\%. The sample was next stirred vigorously for 1 h to obtain an optically clear and homogeneous suspension.

The conductivity of the Laponite suspensions was measured using the Eutech PC 2700 four-probe setup. The zeta potential of the Laponite suspensions was monitored by an electro-acoustic accessory purchased from Dispersion Technology DT 100 \cite{dukhin}. A specially designed electro-acoustic probe (DT-1200) that is cylindrical in shape with a flat metal surface was immersed in the suspensions. The probe, equipped with an ultrasound transducer, generates ultrasound waves at frequency 3 MHz which propagate through the suspension. The diffusive charges of the electric double layer surrounding the charged colloidal particles oscillate in response to the propagating sound waves. The oscillating charges of the particles produce a current which is called the colloidal vibration current (CVI). The resulting CVI and its phase are detected by the gold metal surface of the electro-acoustic probe. The measured CVI data is analyzed by the DT-$1200$ software to compute the $\zeta$ potential. Further details of the instrument and electro-acoustic principles can be found in the literature \cite{dukhin}. For the conductivity measurements, a 4-cell conductivity probe applies an oscillating current signal (at $1$kHz) between two of the electrode cells. The resulting voltage is measured using the other two electrode cells. This four probe technique minimizes electrode polarization effects (i.e the accumulation of charged species on the electrode surface). To ensure the accuracy of the acquired conductivity data, the conductivity probe was calibrated using a series of standard KCl solutions of known conductivities. 

For the DC electric field experiments, two circular brass disks of diameter 10 mm  were used as electrodes connected with thin wires for electrical contacts. A thin Teflon ring of thickness 0.3 mm was kept as a spacer between the two brass plates. This electrode set-up was immersed in the freshly prepared Laponite suspension. A small DC voltage varying between 0.8 V- 3 V (electric field  $E$=2.5 kV/m to 10 kV/m) was applied using an Agilent E3615A DC power supply for 15 minutes. The soft solid samples that were formed on the surface of the positive electrode upon application of the electric field were extracted using a thin Teflon spatula. For AC voltage studies,  an Agilent 33220A function generator was used and the experiments were performed at 1 kHz with a peak to peak voltage of 2 V. 

The rheological behavior of the soft solid samples formed due to the application of electric fields were studied using an Anton Paar MCR 501  stress controlled rheometer in a cone-plate (CP25) geometry (0.048 mm gap, semi vertical angle= $ 5^{\circ} $). To study the frequency dependent response of the storage ($G'$) and loss moduli ($G''$) of the samples, a small oscillatory strain ($\gamma$) of  peak to peak amplitude  0.1 \%  was applied and the angular frequency $ \omega$ was varied between 0.1 - 100 rad/s. Large amplitude oscillatory sweeps ($\gamma=$ 0.1 \%  to 1000\%) were performed at a fixed $ \omega=$  6 rad/s to study the shear melting behavior of the samples. Measurements are repeated five times to ensure the absence of artefacts due to wall slip.

The microstructure of the samples were studied using cryo - field effect scanning electron microscopy (cryo-FESEM). In these experiments, the samples were loaded into a capillary tube (bore size 1 mm) sample holder (Hampton Research, USA) by a syringe needle. Tubes were sealed at both ends and dipped into a liquid nitrogen bath (at -120$^o$C) for 5 min in the cryo-chamber attached to the SEM sample chamber. The sample tube was cut in the middle and sublimated at -90$^\circ$C for 10 min. The sample was coated with a thin layer of platinum at -150$^\circ$C and transferred (by cryotransfer system, PP3000T from Quorum Technologies) to the imaging stage. Images were recorded in the back scattering geometry by applying a voltage of 3 kV to the field emission electron gun.

\section{RESULTS AND DISCUSSIONS}

\subsection{Formation of soft solids}

Optically clear homogeneous aqueous Laponite clay suspensions (concentration $ C_{L} $=3.0 wt \%) were obtained after stirring for 1 h. At this stage, when the aging time or waiting time is $ t_{w} $= 0 h, the suspensions are fluid-like and were labeled as as-prepared samples. As discussed in the previous section, DC-electric fields of various magnitudes are applied for 15 minutes to induce soft solidification in the as-prepared samples. It is found that soft solidification takes place over the metal surface held at the positive potential. The photograph of the gel formed on the surface of the metal electrode is presented in Fig.1 (a) where the rigid transparent soft solid on the surface and edges of the brass plate can be seen. In contrast, the solidification of aqueous Laponite suspensions is not observed upon the application of AC electric fields (Fig.1(b), where only clear watery fluid is seen on the brass plate), even after a very long experimental duration of 1 h. The absence of formation of a soft solid in Fig. 1(b) is understood by considering that the high rate of switching of the AC field is likely to prevent clustering of the Laponite particles, therby inhibiting the process of sample solidification \cite{{sood}, {Nadal}}. We believe that higher peak-to-peak magnitudes of AC fields may be required for electric field induced soft solidification of the samples investigated here. 

\begin{figure}[h]
\centering
\includegraphics[height=4cm]{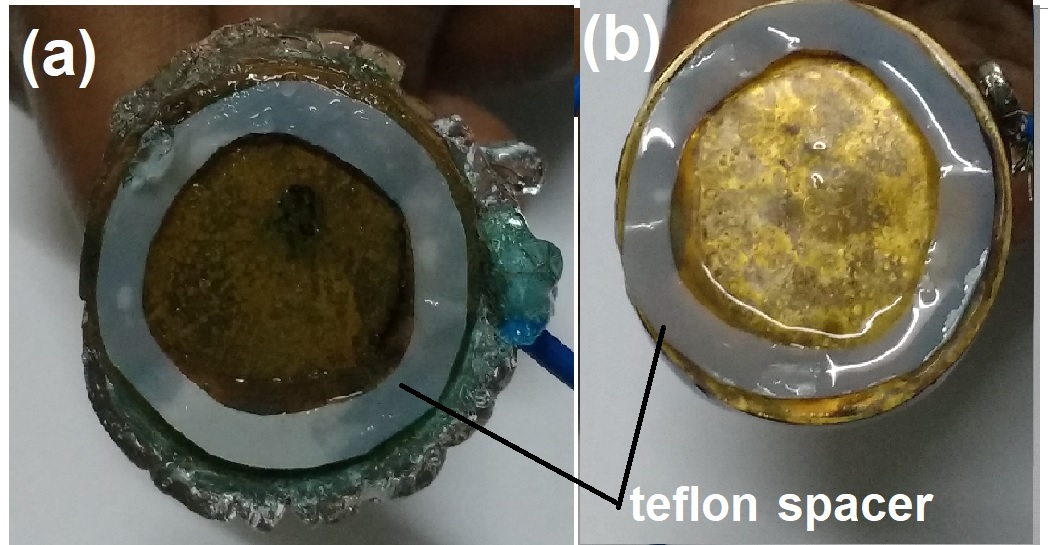}
\caption{(a) Photograph of soft solid formation on the surface of the positive electrode when a DC electric field is applied to a Laponite suspension of concentration $C_{L} $ = 3 wt\%. A thick soft solid supported on the surface and edges of the brass plate can be seen. (b) photograph shows the absence of soft solid in the same sample under an applied AC electric field. In this photo, only watery fluid is seen on the brass plate}
\label{}
\end{figure}

\begin{figure}[h]
\includegraphics [width=9cm]{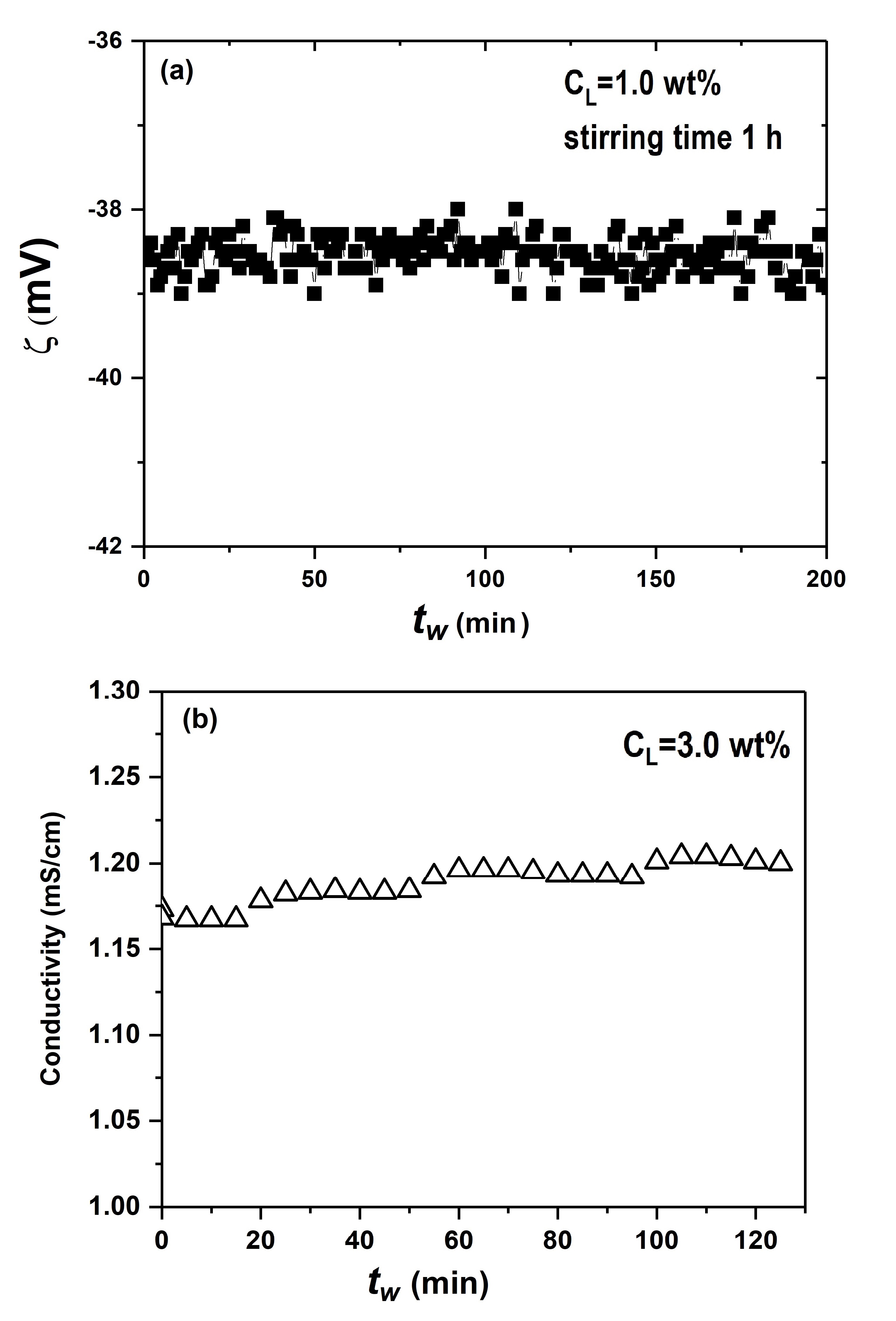}
\caption{(a) Zeta potential measured in a dilute suspension of Laponite of concentration $C_{L} $ = 1 wt\% and (b) conductivity of a Laponite suspension of concentration $C_{L} $ = 3 wt\% with increasing waiting time ($ t_{w}  $).}  
\label{}
\end{figure}

Laponite (XLG) is a high purity synthetic clay of hydrous sodium lithium magnesium silicate and is crystalline in nature. Its chemical composition    (Na$_{+0.7}$[(Si$_{8}$Mg$_{5.5}$Li$_{0.3}$)O$_{20}$(OH)$_{4}$]$^{-0.7}$) matches closely with the natural smectite clay hectorite. Individual Laponite particles are disc-shaped with a thickness of 1 nm and a diameter of 25 nm  \cite{Levitz, Herman, BYK}. The layered structure of each individual Laponite disc consists of an octahedral MgO sheet sandwiched between two tetrahedral silica layers with $ Na^{+} $ ions situated in the intertactoidal space. During sample preparation, the Laponite clay tactoids dispersed in water undergo osmotic-swelling. The swelling process is followed by tactoidal breakup due to the dissociation of $ Na^{+} $ ions in the interlayer galleries of the Laponite tactoids and the diffusion of the latter into the aqueous medium \cite{{ali_langmuir}}. In this delamination process, each Laponite discoid acquires net negative charges over the two flat surfaces (approximately 700 negative electronic charges per particle) and positive charges on the rim (10\% of total negative charge) \cite{{Herman}}. The negative surface charge characteristics of the Laponite particle in aqueous suspension is confirmed from zeta($ \zeta $)-potential measurements in the electro-acoustic set-up. $ \zeta $-potential and conductivity data for the Laponite suspensions are shown in Fig.2. The measured average $ \zeta $-potential is -39.5 mV for $ C_{L} $=1 wt\% Laponite suspension. Conductivity is of the order of 1.16 mS/cm at $ t_{w} $ =0 for a $ C_{L} $=3.0 wt\% Laponite suspension and increases slowly with increasing $ t_{w} $, suggesting the continued dissociation of $ Na^{+} $ ions from unbroken tactoids due to osmosis of the hydrated ions \cite{sahaLagm}.

The $ \zeta $-potential depends on the time for which the suspension has been stirred and shows a small negative value for short stirring times. For the suspensions stirred for 1 h, a constant steady state $ \zeta $-potential (the data is shown in FIG.S1 of Supplementary Information (SI)) is recorded, which signals the completion of the breakup of the clay aggregates. For the sample stirred for only 10 mins, the $\zeta$ potential grows increasingly -ve with increasing $ t_{w} $ and eventually attains a steady state. This confirms the non completion of the process of tactoidal breakup in the samples stirred for shorter times. Colloidal dispersions with aggregates or chained structures are, indeed, reported to show smaller $ \zeta $-potentials \cite{{hunter}, {dukhin}}. 

The strong repulsive interactions between Laponite particles arising out of the net -ve charges on the flat Laponite discoids renders stability to the suspension against van der Waals attraction and gravitational collapse. An equilibrium electric double-layer (DL) (comprising a Stern layer and a diffuse layer) is formed around each particle that can be characterized by a Debye length ($\kappa^{-1}$) whose extent depends on the number of dissociated $ Na^{+} $ ions and  counter ions in the aqueous medium. The DL structure can be modified by altering the $\zeta$- potental which can be accomplished by changing the pH of the medium and by the addition of salts \cite{{hunter}, {dukhin}, {sahaLagm}}. This can lead to the formation of a variety of phases such as attractive colloidal gels, repulsive glasses and structured gels. Furthermore, due to the osmotic pressure induced tactoid exfoliation process discussed earlier, the DL structure in Laponite suspensions evolves spontaneously with increasing sample age ($ t_{w} $) even without the addition of salt or a pH modifier. Eventually, kinetically arrested gel or glass phases are formed due to a physical aging process that has been reported systematically for aqueous Laponite suspensions \cite{{ruzika},{saha},{joshi},{Herman}}. 

In the present experiments, the applied electric field disturbs the DL structure and leads to various linear and nonlinear electrokinetic phenomena \cite{{hunter}, {dukhin}, {Rhodes}}.  It is well known that the application of an electric field can affect the stability of colloids in suspension, with the initiation of a process of displacement of the dielectric colloidal particles (electrophoresis) \cite{{Anderson}}. The electrophoretic velocity ($ U_{0} $) of such a particle far from the electrode is given by $ U_{0}=|(\epsilon \zeta/\eta)E_{0}| $ where $\epsilon$ and $\eta$ are the permittivity and the viscosity of the medium (water) respectively and $E_{0}$ is the magnitude of the field in the absence of particles \cite{{Anderson}}. Three types of mechanisms of particle interactions are expected to exist in an electric field. These are: (i) electrodynamic interaction (due to polarization of the particle by the electric field) which is responsible for the formation of chains of particles in electrorheological (ER) fluids, (ii) electrohydrodynamic processes in which the electric field induces electro-osmotic flow and (iii) a concentration mechanism related to the concentration polarization of the particles. 

The combination of the above electrokinetic mechanisms can explain the soft solid formation of Laponite suspensions in the presence of applied DC electric fields. In the case of Laponite aqueous suspensions, the positive and negative charge distributions give zero net dipole moment and non zero quadrupole moment \cite{tarafdar}.Since the Laponite discoids have a large effective net negative charge, their suspensions have very high conductivities. The electric fields applied in the experiments are therefore expected to change the structure of the electric DL considerably. The $ Na^{+} $ ions migrate towards the negative electrode while the predominantly negatively charged Laponite discs get attracted towards the positive potential due to the net Coulomb force. This gives rise to an electrophoretic drift current of Laponite particles \cite{Anderson}. Since the Laponite particles have a fixed heterogeneous distribution of charges on their surfaces (-ve charges on the surfaces and +ve charges on the edge), they are known to associate into house of cards (HoC) structures \cite{{Laxton}, {aliSM}}. In the present experiments therefore, a process of particle clustering results from electrophoresis and leads to the formation of a soft solid due to the association of the positively charged rims of the Laponite particles to the negatively charged faces. 

In the absence of any external field, it is well known that the electrostatic interactions between Laponite particles build up slowly in the undisturbed suspensions over a period of time and repulsive glasses are formed at appropriate Laponite concentrations. The spontaneous evolution of $ G' $ and $ G'' $ of the suspensions in the absence of an electric field are plotted in the FIG.S2 in the SI. Initially at $ t_{w} $=0 h, the suspensions exhibit liquid-like behavior as only $ G'' $ can be measured ($G'$=0). As the Laponite sample evolves with time due to the exfoliation-driven physical aging process described earlier, the elasticity of the sample develops and increases gradually, and eventually a crossover of $ G' $ and $ G'' $ can be seen. In the case of the external applied steady state electric field, the particles align and move in the direction of the field which results in the onset of rapid solidification as seen in our experiments. It is observed that larger amounts of samples are formed with increasing durations of application of the electric field. Furthermore, the soft solid formation is found to be an irreversible process. Even after the electric field is switched off, the samples are found to remain intact. The irreversible soft solidification of aqueous Laponite suspensions can be attributed to electrokinetic phenomena.

\begin{figure}[h]
\includegraphics [width=9cm]{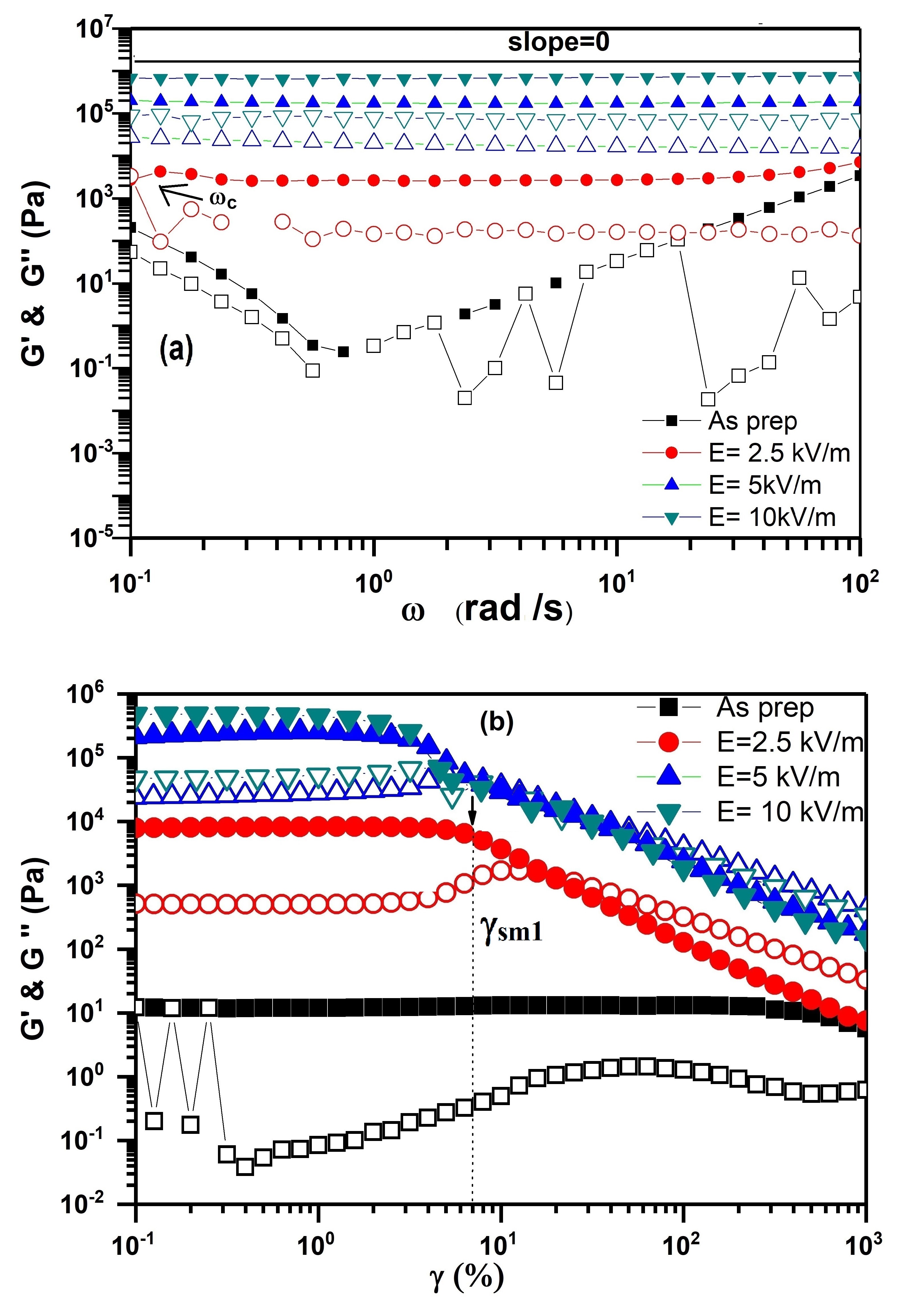}
\caption{ (a) $G' $ and $G''$ vs. angular frequency ($ \omega $ ) of Laponite gels ($ C_{L}$=3 wt\%) formed due to applied DC electric fields. $G'$ is denoted by solid symbols and $G''$ by open symbols. A solid line of slope zero is drawn as reference and data at $E$> 2.5 kV/m are seen to approximately follow the prediction of the SGR model. (b) Storage ($G'$) and loss ($G''$) modulus vs. applied oscillatory strain ($ \gamma $) amplitude. }  	
\label{}
\end{figure}

\subsection{Rheological characteristics}

The soft solids that are formed due to the application of electric fields of various strengths ($E$) are extracted carefully and subjected to rheological studies in a rheometer. The results are depicted in Fig.3. In the absence of an electric field, the suspensions are fluid-like and show a very small elastic modulus (solid squares in Fig. 3). The storage ($G'$) and loss ($G''$) moduli are seen to increase by several orders of magnitude when $ E $ is increased from 2.5 to 10 kV/m when compared to the as-prepared sample (circles, up-triangles and down-triangles in Fig.3 correspond respectively to $E$ =2.5, 5 and 10 kV/m). For all $E$>0, the $G' $ value is consistently greater than $G''$ in all the samples, thereby confirming the viscoelastic solid-like nature of the samples. The evolutions of $G' $ and $G''$ in these experiments indicate the onset of a liquid-to-solid transition in the presence of the applied steady electric field. It is also noted that the $G'$ and $ G''$ of these samples are frequency independent and develop plateaus at higher $E$ values ($E$ = 5 and 10 kV/m). This is reminiscent of the rheology predicted for soft solids by the soft glassy rheology (SGR) model, where the frequency responses of $G'(\omega) $ and $ G''(\omega) $ in the linear viscoelastic regime are seen to exhibit power-law dependences of the type $G' (\omega)$, $G''(\omega) =A\omega^{x-1} $ where $ A $ is a constant and $ x $ is an exponent  \cite{{Sollich}}. In this model, $x$ is a noise temperature and $x$=1 signals the onset of the glass transition. In Fig. 3(a), we have plotted a solid line for the equation $x$=1 (slope=0). We note that both $G'(\omega)$ and $G''(\omega)$ measured for the samples at $ E $>0 are approximately parallel to the $x$=1 line, thereby confirming the soft glassy rheological response of these samples. For low values of the applied electric field, $ G'' $ increases weakly with decreasing $ \omega $ and eventually shows a crossover with  $G'$. The crossover point $ \omega_{c} $ is identified as the relaxation frequency of the samples. For higher  $E$, strong structured samples are formed. These samples are expected to have very slow dynamics which precludes the measurements of their long relaxation times within the experimental window. Due to the strengthening of the soft solid with increasing electric field, the crossover of $G'$ and $G''$ cannot be seen in the data acquired at $ E $>5 kV/m. The data plotted in Fig.3(a) therefore clearly demonstrates the remarkable acceleration of the dynamical slowing down process in the samples formed by applying high $E$ values ($E$>5 kV/m).
	
\begin{figure}[h]
\includegraphics [width=9cm]{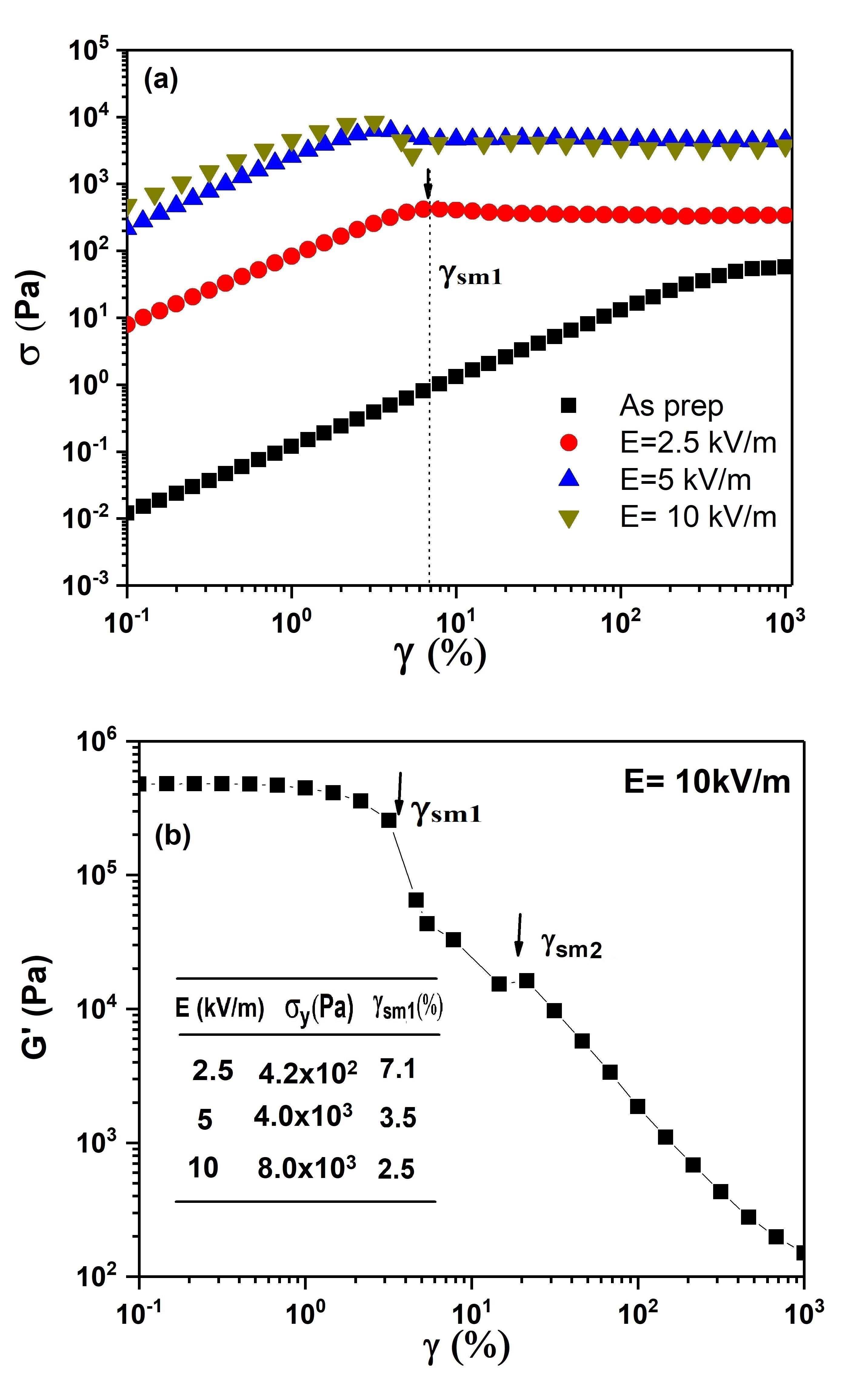}
\caption{(a) Shear stress vs. shear strain curves for a soft solid of Laponite ($C_{L} $ = 3 wt\%) and (b) the two-step yielding response of a Laponite soft solid formed at $E$=10 kV/m . The table in the inset lists the yield stress $ \sigma_{y} $ and the first yield strain value $ \gamma_{sm1} $ for the soft solids formed at different $E$ values.}
\label{}
\end{figure}
	
The shear melting of the samples is studied by strain amplitude sweep measurements and is shown in Fig. 3(b) for different $E$ values. The plateau region in the $G'$ and  $G''$ data at low $\gamma$ values corresponds to the linear viscoelastic regime. It can be seen that with increasing E, the linear viscoelastic regime shifts to lower strain amplitudes $ \gamma $.  As the strain is increased beyond the linear viscoelastic regime for that particular sample, the soft solids shear melt at a strain $ \gamma_{sm1} $ which is determined as the value of the strain amplitude ($\gamma$) at which the drop in the elastic modulus is approximately 5\% \cite{Laurati}. It is seen that $ \gamma_{sm1} $ decreases with increasing $E$. This is further confirmed from the plots of the oscillatory shear stress vs. $ \gamma $ shown in Fig. 4(a). We observe from this figure that all the samples show Herschel-Bulkley like yield stress and shear thinning behaviors, which is also predicted by the SGR model for soft colloidal materials \cite{{Chin},{Sollich},{goodwin}}. The yield stress $ \sigma_{y} $ is determined at the onset of shear melting and corresponds to the value of $ \sigma $ at which the linear plot  deviates by 5 \% from the prediction of Hookes law. It is seen that $ \sigma_{y} $  increases with increasing  $E$ (table in the inset of Fig.4(b)). Apart from $ \sigma_{y} $, $ \gamma_{sm1} $ values are also listed in the table displayed in the inset of Fig.4(b). It is noted that the samples formed in the presence of higher electric fields shear melt at smaller strain amplitudes. This indicates the mechanically fragile behavior of the samples \cite{{Chin}} that can be associated with the rapid solidification process in the presence of the DC electric field. The small values of $ \gamma_{sm1} $ can be attributed to the soft screened electrostatic interactions among the particles. Moreover, the samples subjected to higher $E$ show a clear two-step shear melting behavior (seen clearly in Fig.4(b) for $E$=10 kV/m) within a narrow window of strain  $ \gamma $ . The first step (denoted as $ \gamma_{sm1}$ in Figure 4(a)) can be linked to the breakup of the clustered network structure, i.e due to the breakdown of the inter-linked structures of clusters or chains of particles of the samples. The second step (denoted as $ \gamma_{sm2}$ in Fig.4(b), indicated by a deviation from a power law behavior at $ \gamma$>$ \gamma_{sm1}$ ), can be attributed to the breakup of the individual clusters or chains \cite{{petekidis}}.

\begin{figure*}[!t]
			
\includegraphics [width=\textwidth]{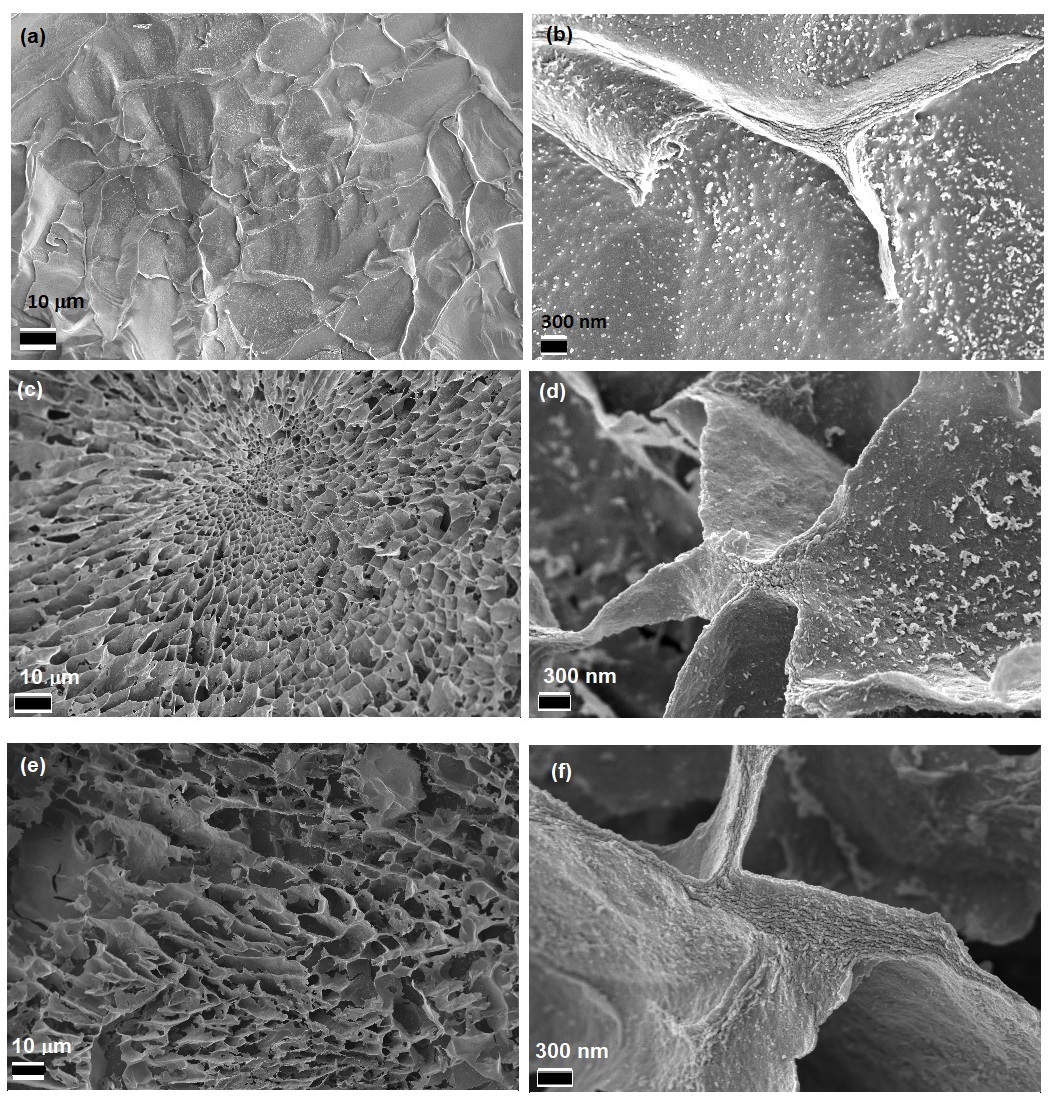}
\caption{Cryo-SEM micrographs of Laponite gels ($ C_{L} $= 3 wt \%) formed at various electric fields (a) $E$= 2.5 kV/m, (c) $E$=5 kV/m and (e ) $E$=10 kV/m. The scale bar corresponds to 10 $\mu$m. The right panel images (b),(d) and (f) are magnified version of the images displayed in (a), (c) and (e) respectively. The scale bars in (b),(d) and (f) correspond to 300 nm.} 	
\label{}
			
\end{figure*}

\subsubsection{Comparison with electrorheological (ER) fluids}

	The rheological behavior of the electric field induced Laponite soft solids discussed above is reminiscent of positive ER like behavior i.e., dramatic increase in the rheological properties (yield stress, storage modulus etc.) with increase in the applied electric field \cite{YGKo}.  However, unlike conventional ER fluids, the transformation into the soft solid reported here is irreversible.  Furthermore, by applying the electric field in the rheometer parallel-plate geometry cell (rheo-dielectric cell), we observe that the soft solidification or gelation occurs over the surface of the positive electrode (the photograph is presented in Fig.1 (a) and in SI as Fig. S3). This gelation phenomenon is not homogeneous throughout the gap of the sample cell. As a result of such inhomogeneous sample deposition, it is not possible to conduct in situ rheological studies. The positive ER property of Laponite particles dispersed in an insulating oil can be found in the literature \cite{parmar}.
	
	Furthermore, we have verified that these observations are reminiscent of the behavior of negative ER fluids in which phase separation (electromigration) was noticed due to the migration of the ER particles toward one or both of the electrodes due to electrophoresis under an applied electric field \cite{Boissy, YGKo,MMRamos,YGKo1,HHuang}. The negative ER effect is known to produce the opposite effect of positive ER behavior, i.e the application of an electric field decreases the rheological moduli due to the destruction of structures \cite{Boissy, MMRamos}. Suspensions of elongated goethite ($\beta$-FeOOH) particles in silicone oils with varying amounts of silica nanoparticles \cite{MMRamos} and alginic acid and alginate salts dispersed in silicone oils show negative ER behavior and the electromigration of particles towards both the electrodes  \cite{YGKo1}. In contrast, in the present study, we observe the electrophoretic migration of largely negatively charged Laponite nanodiscs towards only the positive electrode and the subsequent build-up of structures which lead to the formation of the soft solid. The confocal microscopy imaging of these studies, carried out separately using an ITO coated transparent electrode cell filled with Laponite suspension, yielded data of very poor quality due to the nanoscopic dimensions of the Laponite particles and the optically clear nature of their suspensions. No bridging of the particles or formation of pearl chains, as expected in ER fluids, were observed between the two electrodes. Our study verifies the non-ER fluid behavior of aqueous Laponite suspensions under applied DC electric fields.
		
\subsection{Microstructural features}

Finally, we confirm these rheological characteristics directly by visualizing the microstructures of the samples formed by applying DC electric fields in a detailed cryogenic-SEM study. Cryo-SEM images of the electric field induced Laponite soft solids are presented in Fig. 5(a-f) for  $E$=2.5 to 10 kV/m. A honeycomb-like network microstructure is seen in images acquired at lower magnifications ((Fig. 5(a), 5(c) and 5(e) correspond respectively to $E$=2.5, 5 and 10 kV/m and the scale bar  corresponds to 10 $\mu$m) in the left panel of Fig. 5. At small electric field strengths ($E$=2.5 kV/m), the sample in Fig.5(a) shows an open network like structure, i.e with reduced connectivity of the strings or chains of Laponite particles, which results in the small storage modulus and lower yield stress as reported earlier. As the magnitude of the electric field increases ($E$ >2.5 kV/m), a dense percolated network structure (with increased connectivity of the particle strings), spanning the entire space is clearly observed (Fig. 5(c) at $E$= 5 kV/m, Fig. 5(e) at 10 kV/m)). These data explain the observed increase in the storage moduli and yield stresses of the samples reported earlier in the rheological measurements. The magnified images of the network structures of the samples (shown on the right panel of Fig.5, Fig.5(b), 5(d) and 5(f) for $E$= 2.5, 5 and 10 kV/m respectively with the scale bar corresponding to 300 nm) confirm that the walls of the sample network are formed by strings of Laponite particles that are stacked in layers. The possible preferential particle configurations of anisotropic charged clay plates dispersed in an aqueous medium were inspected in Monte Carlo simulation studies at various particle volume fractions and salt concentrations \cite{Delhorme}. With increase in salt concentration in the suspension medium, a sequence of configurations: overlapping coins (OC), house of cards (HoC) and stacked plates, were observed. The gel structures uncovered here from direct cryo-SEM imaging are found to be mixed configurations of HoC and OCs. It is to be noted that while the simulated gels considered the association of individual particles, the experimental gels are formed due to the percolation of small stacks of particles. The honeycomb microstructural network and the network branches comprising stacks of Laponite particle strings that are seen in these cryo-SEM images can give rise to the two-step yielding behavior of the samples observed in Fig. 4(b). It is expected that the first yielding step occurs due to the rupture of the network branches of the honeycomb structure, followed by the breakup of the particle stacks in the individual network branches. This microstructural study clearly demonstrates the formation of gel-like structures due to the application of DC electric fields and qualitatively corroborates the rheological behavior of the electric field induced gels discussed earlier. 
	
\subsection{Electric field induced gels vs. conventional gels}	

In general, colloidal gels are formed by the addition of salt to the suspensions. Addition of salt minimizes the long range electrostatic repulsive interaction among the charged Laponite particles due to screening of the charges by oppositely charged ions. This reduces the Debye length or effective particle radius. The reduction in the repulsive electrostatic forces leads to a weak van der Waals attraction between the particles, and the suspensions are seen to transform to a gel network. However, at higher salt concentration, particles flocculate and phase separate \cite{ali, ali2,Delhorme}. The salt-induced gelation of Laponite suspensions and their properties have been reported in detail \cite{sahin, vkesh}. The Laponite gels obtained after 12 days of aging with 7mM concentration of the salt have storage moduli $ G'$ < 10$^{3}$ Pa \cite{sahin, vkesh}. In contrast, in the present study, DC electric field induced gels are formed much more rapidly and are mechanically much stronger than the salt induced gels. The applied electric field results in the drift of the charged species towards the oppositely charged electrodes. This increases the Debye length associated with each Laponite particle (increasing thereby the effective particle size) as the screening ions are displaced by the electric field. The unscreened charged Laponite particles arrange in clusters of OCs and HoCs (as confirmed from cryo-SEM studies) due to edge to face Coulombic attraction. This process leads to the formation of mechanically strong gels having storage moduli three orders of magnitude higher than those of the salt induced gels formed after several days. Therefore, the present study suggests that strong gels can be prepared rapidly by applying appropriate magnitudes of DC electric fields to aqueous suspensions of charged colloidal particles.
		
\section{Conclusions}
In this work, the soft solidification of aqueous Laponite clay suspensions, induced by DC electric fields, is studied by varying the electric field strength.  The $ \zeta $-potential of the Laponite particles dispersed in water is seen to be negative due to the dissociation of $ Na^{+} $ ions from the intertactoidal spaces of the Laponite stacks. These Laponite suspensions are seen to exhibit large electrical conductivities. It is observed that soft solidification occurs over the surface of the positive electrode. The speed of solidification increases with increasing magnitude of the electric field. The phenomenon of soft solidification at the positive electrode is attributed to the large effective negative surface charge on the Laponite discoid, the net positive charge on the rim and to the resultant electrokinetic phenomena. In the experiments when an AC electric field (at frequency of 1 kHz) is applied, no solidification is observed due to the fast switching of the electric field that prevents the clustering of the Laponite particles. 
	
The mechanical behavior of the samples is probed in a rheometer and a fluid to soft solid transition at various electric field strengths is clearly observed. The storage moduli and yield stresses of the samples increase with the increase in the electric field strength. The small yield strain values measured for the samples and its decrease with increasing electric field strengths confirm the mechanically fragile nature of the soft solids. In the presence of small oscillatory strains, the elastic modulus shows a frequency independent response, while the loss modulus is weakly dependent on frequency. Samples formed at higher electric fields  exhibit two-step shear melting behavior which is attributed to the rupture of the network of particle clusters/chains, followed by the breakup of the individual clusters/chains. The microstructural characteristics of the soft solids are studied using cryo-scanning electron microscope and clearly demonstrate the formation of  honeycomb like network structures upon increasing the electric field strength. The network connectivity of the samples increases with increasing electric field strength, thereby increasing their storage moduli and yield stresses. The magnified views of the network walls delineate the presence of strings of Laponite particles stacked into layers and corroborate the two step yielding behavior of the samples reported in our rheological experiments. Furthermore, the electric field induced gels are much stronger than the conventional salt induced gels due to unscreened charged particles participating in gelation in the former case.
	
The present study contributes a new complex strong gel-like phase to the rich phase diagram of nonequilibrium Laponite suspensions undergoing physical aging in the presence of an externally applied electric field. The study also demonstrates that control of the rheological properties of Laponite colloidal suspensions and their mechanical strengths can easily be achieved by simply varying the DC electric field strengths. We believe that the present study will encourage researchers to simulate the soft solidification of Laponite colloids under external stimuli. Such efforts can lead to the eventual development of smart fluid-like responses in colloidal Laponite suspensions subjected to DC electric fields.

\section*{{Acknowledgments}}

We acknowledge A.Dhasan and Y.M. Yatheendran for their help in recording cryo-SEM images.


\end{document}